%% file: Doppler_space_key.tex
\documentclass[letterpaper, 10 pt, conference, right=0.6in]{IEEEtran}
\IEEEoverridecommandlockouts
\usepackage{cite}
\usepackage{amsmath,amssymb,amsfonts}
\usepackage{amsthm}
\usepackage{algorithmic}
\usepackage{graphicx}
\usepackage{textcomp}
\usepackage{float}
\usepackage{xcolor}
\usepackage{multirow}
\usepackage{subfigure}
\usepackage{adjustbox}
\usepackage{stfloats}
\def\BibTeX{{\rm B\kern-.05em{\sc i\kern-.025em b}\kern-.08em T\kern-.1667em\lower.7ex\hbox{E}\kern-.125emX}}
\begin{document}

\title{\LARGE Securing the Inter-Spacecraft Links: \\ Doppler Frequency Shift based Physical Layer Key Generation\\
{}
}

\author{ \IEEEauthorblockN{ Ozan Alp Topal\IEEEauthorrefmark{2}, Gunes Karabulut Kurt\IEEEauthorrefmark{2}, Halim Yanikomeroglu\IEEEauthorrefmark{3}}
\IEEEauthorblockA{\IEEEauthorrefmark{2} \textit{Department of Electronics and Communication Engineering}, \textit{Istanbul Technical University}, Istanbul, 34469 Turkey  \\
\IEEEauthorrefmark{3}  \textit{Department of Systems and Computer Engineering}, \textit{Carleton University}, Ottawa, ON K1S 5B6, Canada	\\
E-mail:	\{topalo, gkurt\}@itu.edu.tr, halim@sce.carleton.ca}
}

\maketitle

\begin{abstract}
We propose a novel physical layer secret key generation method for the inter-spacecraft communication links. By exploiting the Doppler frequency shifts of the reciprocal spacecraft links as a unique secrecy source, spacecrafts aim to obtain identical secret keys from their individual observations. We obtain theoretical expressions for the key disagreement rate (KDR). Using generalized Gauss-Laguerre quadrature, we derive closed form expressions for the KDR. Through numerical studies, the tightness of the provided approximations are shown. Both the theoretical and numerical results demonstrate the validity and the practicality of the presented physical layer key generation procedure considering the security of the communication links of spacecrafts. 
\end{abstract}


\section{Introduction}
 {\color{black}As the commercial opportunities of space exploration flourish, the space networks become the next frontier in wireless communication. Thanks to the recent advances in rocket launch platforms to launch new spacecraft to the space, availability of dedicated frequency spectrum, the availability of lower complexity and smaller devices have lowered the cost of spacecraft supported services including space travels. Today mostly the small-space satellites are launched into low-orbital space by top-tier companies like SpaceX, Google, Facebook, Virgin Galactic\cite{first_ref}. 
 
 However, exploration of the outer space is expected to become the near future competition for the countries and the private companies. For example, with the Artemis program, NASA targets to land the first human on the Moon by 2024\cite{NASA}. By the end of the decade, NASA is planning to form sustainable space missions that eventually sending astronauts to Mars. In order to join the space exploration race, these spacecrafts require innovative technologies that provide a lower cost of production, launch, and maintenance. This requirement limits the sensing and the communication capabilities predicting their limited size, weight, and power. Considering these characteristics, the spacecrafts that will be utilized in the future space missions act as space cyber-physical networks, where the control and connectivity of many low-cost and software-enabled controllable devices are the main priority.}

 
 \begin{figure}[t]
 	\centering
 	\includegraphics[width=\linewidth]{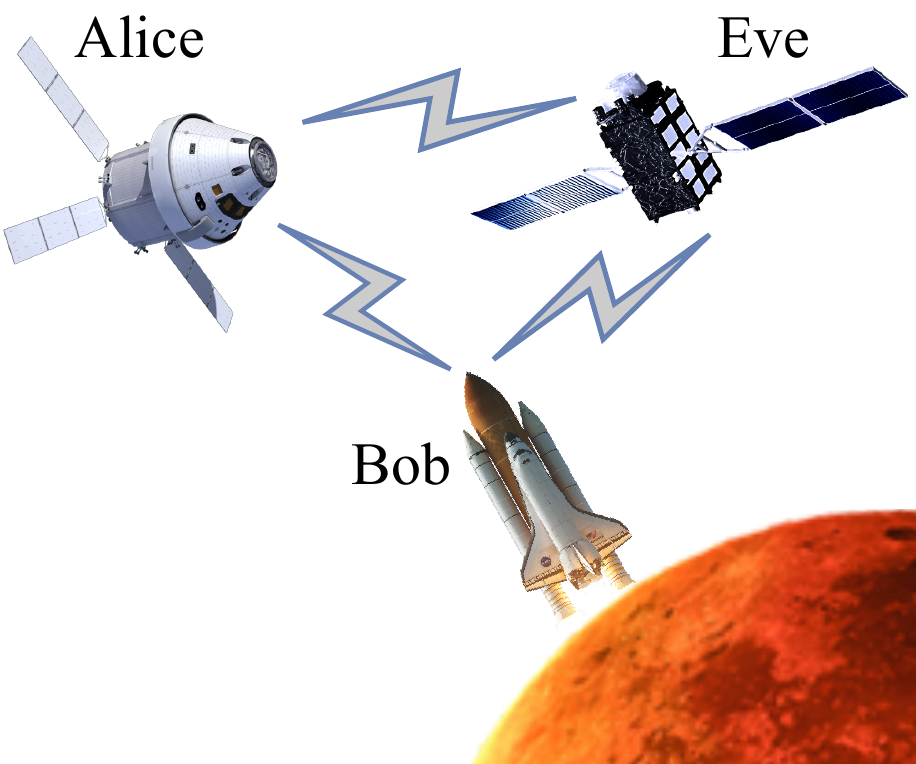}
 	\caption{An illustration of an ISL. Alice, Bob and Eve represent three different {\color{black}spacecraft}s. Alice and Bob try to establish a secure communication line, while Eve tries to eavesdrop their communication.}
 	\label{fig:LEO}
 \end{figure}

 
 {\color{black}The open nature of the wireless communication channel creates security breaches for the space networks similar to the wireless cyber-physical systems. Especially considering the critical research and exploration information harnessed by the space mission, the security breach of an inter-spacecraft link (ISL) would create the loss of highly profitable information. Considering the high operational costs and technical difficulties of restarting a space mission}, the security becomes a non-negotiable concept in space networks. Conventionally, the security has enabled by cryptographic methods, where the message is encrypted with a secret key at the transmitter and decrypted with the same key at the receiver. This secret key is obtained by solving a series of complex mathematical problems that require a high computation power, constituting a challenge for the {\color{black} envisioned low-complexity spacecrafts}.

Physical layer key generation exploits the unique characteristics of the physical layer link between two physically separated nodes to extract an identical secret key. {\color{black}  Any other node would not be capable to generate the same key.} Previously, received signal strengths (RSSs) \cite{RSS}, channel gains \cite{Topal}, channel phases or spatial correlations of the channel fadings \cite{Topal_MIMO} are utilized as a common secrecy source in physical layer key generation. Considering the ISLs, continuous randomness cannot be attained by both RSS and  channel fading based methods since the channel attenuation is resulted from path loss that is predictable. Based on the following observation, we propose a fresh perspective for the physical layer key generation in ISLs.
   
 \noindent\textbf{Observation 1:}\textit{ Let us denote the relative velocity of {\color{black}spacecraft} $m$ with respect to the {\color{black}spacecraft} $n$ by $v_{mn}$. The corresponding Doppler frequency can be calculated by $\omega_{mn}=v_{mn}f_0/c$, where $f_0$ is the carrier frequency of the transmitted electromagnetic wave and $c$ is the speed of the light \cite{speed}.  The Doppler frequency of the reciprocal links can be uniquely described with $\omega_{mn}=-\omega_{nm}$, while the Doppler frequency of received signal from any other physically disjoint node $t$ will be different than other ISL's $\omega_{mt} \neq \omega_{mn} \neq \omega_{nt}$.} 
 
 From Observation 1, we can state that the Doppler frequency is a shared secret between reciprocal nodes, while other nodes cannot obtain this secret. Motivated by this observation, we will present a novel Doppler frequency based secret key generation procedure. The proposed method is based on collecting pilot signals transmitted from two separated nodes and utilizing the nominal power spectral density samples (NPSDSs) that will be identical for symmetric Doppler frequencies. {\color{black} As described in \cite{Spectrum}, one of the key requirements for a {\color{black}spacecraft} is estimating and overcoming the Doppler frequency shift. We exploit this process to generate a common secret between two distant {\color{black}spacecrafts} without introducing any additional complexity to the system.} The main contributions of this paper can be listed as 
\begin{itemize}
\item We propose a security mechanism specifically designed for the ISLs for the first time in the literature.  
\item  A novel physical layer key generation method that exploits identical observations based on the Doppler frequency shifts at two distant {\color{black}spacecrafts} is proposed. {\color{black} Considering continuously changing mobility of {\color{black} the spacecrafts}}, the proposed method ensures that the generated key cannot be duplicated by any other physically disjoint node. 
\item Theoretical key disagreement rate (KDR) expressions are derived considering the estimation errors at the distant nodes. Tight approximations to KDR expressions are obtained by using Marcum-Q functions and generalized Gauss-Laguerre quadrature (GLQ).
\item Numerical results are given to verify the theoretical expressions. The tightness of the provided approximations are shown.
\end{itemize}
 
In the following, we provide related work in physical layer key generation. In Section II, we explain the proposed secret key generation procedure. In Section III, we provide theoretical analysis on the key disagreement rate of the proposed method. In Section IV, we provide numerical analysis. In Section V, we conclude the paper and provide the future work directions.

\subsection{Related Works}
In \cite{hanzo}, the authors provide fundamental steps of the physical layer key generation, and compare the existing channel-based key generation methods. Alternatively, the authors of \cite{phase} introduce a key generation mechanism based on channel phase under narrowband fading assumption. In \cite{static}, the authors utilize a relay node in secret key generation to overcome the static environment characteristics. In \cite{underwater}, the authors propose a secret key generation system for underwater communication channels. In addition to different key generation schemes, another major contribution is increasing the efficiency of the proposed key generation mechanisms. The authors of \cite{PCA} make use of principle component analysis to decrease the key disagreement rates at the nodes. In \cite{Topal_wavelet}, the authors utilize a wavelet-based pre-processing to eliminate the dissimilarities of the channel observations.  

{\color{black}As described in the Figure \ref{fig:LEO}, we consider any three ISLs of three {\color{black} distinct spacecrafts}.} In the considered scenario, Alice and Bob exploit the symmetric measurements of the Doppler frequencies as a secrecy source for the first time in the literature. Since, the relative velocity of the {\color{black}spacecraft}s is not static over time, the generated key can be updated in each time periods. 

\section{Secret Key Generation Procedure}
\begin{figure}[t]
	\centering
	\includegraphics[width=\linewidth]{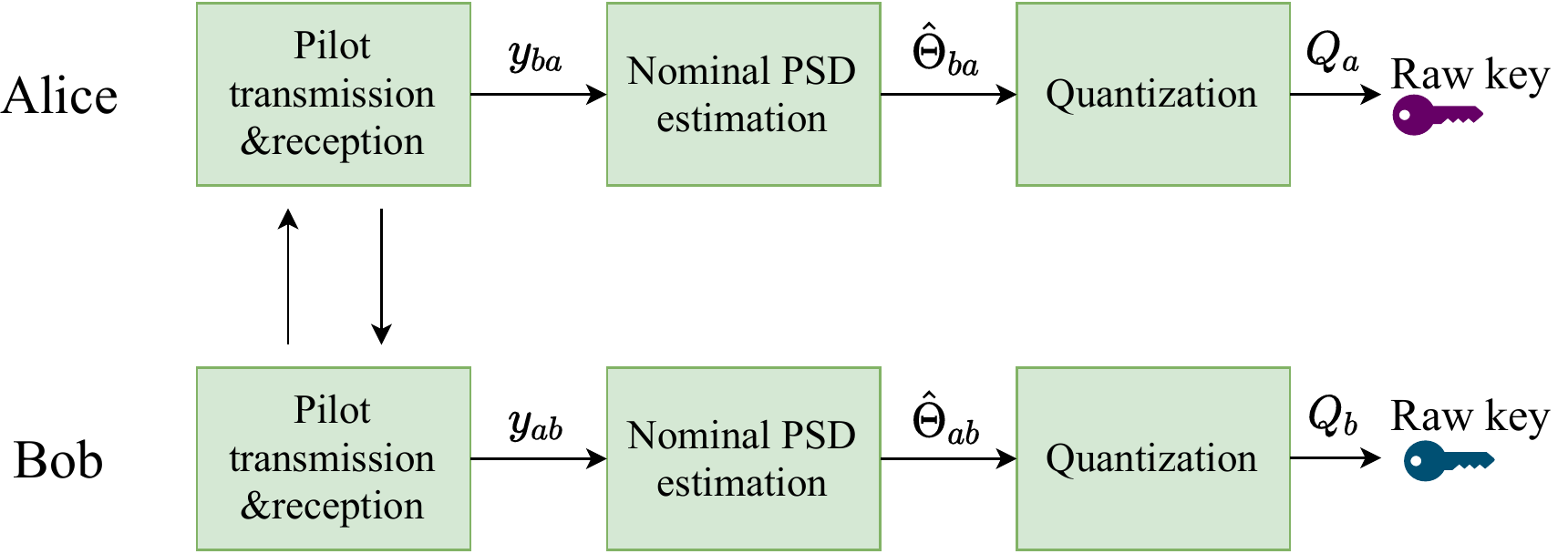}
	\caption{An illustration of secret key generation procedure.}
	\label{fig:system_model}
\end{figure}
\subsection{Pilot Transmission \& Reception}
The secret key generation procedure is described in the Figure \ref{fig:system_model}.
 As a first step, Alice and Bob respectively transmit $N$ pilot symbols $\mathbf{x}= \left[x(1), x(2), \ldots, x(N) \right]$ in a time division duplexing (TDD) fashion. The considered {\color{black}spacecraft}s Alice, Bob and Eve are respectively be denoted by $\{a,b,e\}$.   Considering the ISLs, the baseband representation of the received symbols at the {\color{black}spacecraft} $k$ from the {\color{black}spacecraft} $j$, $\mathbf{y_{jk}}=\left[y_{jk}(1), y_{jk}(2), \ldots, y_{jk}(N) \right]$ can be modeled as \
\begin{equation}
y_{jk}(i)=x_{jk}(i)+\kappa_{jk}(i),
\end{equation}
where $i\in\{1,\ldots,N\}$, $x_{jk}(i)=\zeta x(i)e^{j\omega_{jk}iT}$, $\zeta=\frac{1}{d^{PL}_{jk}}$ and $j\in\{a,b\}$, $k\in\{a,b,e\}$ and $j\neq k$ \cite{LEOsat}. $\zeta$ denotes the path loss attenuation, where $d_{jk}$ is the distance between spacecraft $j$ and $k$, and $PL$ is the path loss exponent. $\omega_{jk}$ denotes the Doppler frequency shift for the link $j-k$, and $T$ is the symbol period.  $\kappa_{jk}(i)\sim \mathcal{CN}(0,\sigma_{k}^2)$ denotes the additive white Gaussian noise at the receiver $k$, where $\mathcal{CN}(0,\sigma^2)$ denotes the i.i.d. complex normal distribution with zero mean and $\sigma^2$ variance. Considering the similar quality of the radio equipments at each {\color{black}spacecraft}, we assume that the noise variance is equal for each receiver as $\sigma_{k}^2= \sigma^2$. 

The information contained in $\mathbf{y_{jk}}$ is fully present in its discrete Fourier transform: 
\begin{equation}
Y_{jk}(i)=X_{jk}(i)+K_{jk}(i),
\end{equation}
where $\mathbf{Y_{jk}}=\mathcal{F}\{y_{jk}\}=\left[Y_{jk}(1), Y_{jk}(2), \ldots, Y_{jk}(N) \right]$, $\mathbf{X_{jk}}=\mathcal{F}\{x_{jk}\}=\left[X_{jk}(1), X_{jk}(2), \ldots, X_{jk}(N) \right]$ and $\mathbf{K_{jk}}=\mathcal{F}\{\kappa_{jk}\}=\left[K_{jk}(1), K_{jk}(2), \ldots, K_{jk}(N) \right]$. Since Gaussian processes are invariant against Fourier transform, the signal spectrum $\mathbf{X_{jk}}$, and the noise spectrum $\mathbf{K_{jk}}$, are also complex Gaussian, zero-mean, and orthogonal processes. The spectral samples ${Y_{jk}(i)}$ are mutually uncorrelated because of the assumed stationarity of $X_{jk}(i)$.

Note that, as stated in \cite{crbsystem}, the phase of $\mathbf{Y_{jk}}$ carries no information about the Doppler frequency, since $\mathbf{Y_{jk}}$ has been modeled as a stochastic process with the aforementioned properties. Hence, it is sufficient to consider the power spectrum of the received data as 
\vspace{-1mm}
\begin{equation}
\begin{aligned}
\mathbf{S_{jk}} &= \left[S_{jk}(1), S_{jk}(2), \ldots, S_{jk}(N)\right] \\ 
&= \left[|Y_{jk}(1)|^2, |Y_{jk}(2)|^2, \ldots, |Y_{jk}(N)|^2\right].
\end{aligned}
\end{equation}

Since $Y_{jk}(i)$ is a complex Gaussian process, the probability density function of each sample ${S_{jk}}(i)$ under the condition of a particular Doppler frequency $\omega_{jk}$ is given by the exponential distribution \cite{crbsystem}:
\vspace{-2mm}
\begin{equation}
\rho(S_{jk}(i);\omega_{jk})= \frac{1}{\Theta_{jk}(i)}\text{exp}\left(-\frac{S_{jk}(i)}{\Theta_{jk}(i)}\right), 
\end{equation}
where $\Theta_{jk}(i)$ denotes the NPSDS, and can be obtained by  
\begin{equation}
\begin{aligned}
\Theta_{jk}(i)= \mathbb{E}\{{S_{jk}}(i)\} &= \mathbb{E}\{|X_{jk}(i)+K_{jk}(i)|^2\} \\
&=\mathbb{E}\{|X_{jk}(i)|^2\}+\mathbb{E}\{|K_{jk}(i)|^2\}
\end{aligned},
\end{equation}
where $A^x_{jk}(i\Delta f-\omega_{jk})=\mathbb{E}\{|X_{jk}(i)|^2\}$, $A^n_{jk}=\mathbb{E}\{|K_{jk}(i)|^2\}$ and $\mathbb{E}\{\cdot\}$ denotes the expectation operator. Note that $A^x_{jk}(f)$ is the a priori known nominal power spectral density of the signal; $\Delta f$ is the frequency sampling interval; and $\omega_{jk}$ is the Doppler frequency shift. Considering $A^x_{jk}(f)$ is periodic with period $\Delta f$, and $A^n_{jk}$ is a constant, we can deduce that $\Theta_{jk}(i)$ is also periodic with $\Delta f$, and consequently we can drop $i$ and denote the NPSDS as $$\Theta_{jk}=\Theta_{jk}(i), \forall i .$$


\textbf{Proposition 1:} The NPSDSs for reciprocal links will be equal to \begin{equation}
\Theta_{jk}=\Theta_{kj},
\end{equation}

while any other {\color{black}spacecraft} than $j$ and $k$ would observe different values. 
\vspace{-2mm}
\begin{proof}
We can proof the Proposition 1 in two steps:
\begin{enumerate}
\item The NPSDS follows the symmetric relation as $A^x_{jk}(i\Delta f-\omega_{jk})= A^x_{jk}(i\Delta f+\omega_{jk})$ \cite{proakis}. As stated in Observation 1, $\omega_{jk}=-\omega_{kj}$. Considering the same pilot sequence is transmitted from both {\color{black}spacecrafts}, $A^x_{jk}(f)=A^x_{kj}(f)$.  Consequently, we can state that  $ A^x_{jk}(i\Delta f+\omega_{jk})= A^x_{kj}(i\Delta f+\omega_{jk})$.
\item As mentioned above, the variance of the thermal noise at each receiver is assumed to be equal. Therefore, $A^n_{jk}=A^n_{kj}$ and $$
\underset{\Theta_{jk}}{\underbrace{A^x_{jk}(i\Delta f+\omega_{jk})+A^n_{jk}}} =\underset{\Theta_{kj}}{\underbrace{A^x_{kj}(i\Delta f+\omega_{kj})+A^n_{kj}} } 
 $$
\end{enumerate}
\vspace{-2mm}
\end{proof}

As indicated in Proposition 1, the ideal NPSDS at Alice and Bob would be equal. Without explicitly calculating the Doppler frequency, {\color{black}spacecrafts} may individually estimate the NPSDS from their observation sequences, and they can exploit this value as a shared secret for key generation mechanism. In the following, we describe the estimation of the NPSDS process at the {\color{black}spacecraft}s.

\subsection{Nominal PSD Estimation}
Let us denote the estimated NPSDS at receiving node $k$ as $\hat{\Theta}_{jk}$. The maximum likelihood (ML) estimation of the parameter ${\Theta}_{jk}$ can be given as:

\begin{equation}
\begin{aligned}
\hat{\Theta}_{jk} &= \max_{{\Theta}_{jk}}\left\{ \prod\limits_{i=1}^{N} \frac{1}{{\Theta}_{jk}}\text{exp}\left(-\frac{S_{jk}(i)}{\Theta_{jk}}\right) \right\} \\
&=\max_{{\Theta}_{jk}}\left\{\frac{1}{{\Theta}^N_{jk}}\text{exp}\left(-\frac{\sum\limits_{i=1}^{N}S_{jk}(i)}{\Theta_{jk}}\right) \right\}. 
\end{aligned}
\end{equation}

The log-likelihood function for the estimation problem can be given as:
\begin{equation}
\mathcal{L}(S_{jk}(i);{\Theta}_{jk})= -N\ln({\Theta}_{jk})-\left(\frac{\sum\limits_{i=1}^{N}S_{jk}(i)}{\Theta_{jk}}\right).
\end{equation}

\begin{figure}[t]
	\centering
	\includegraphics[width=\linewidth]{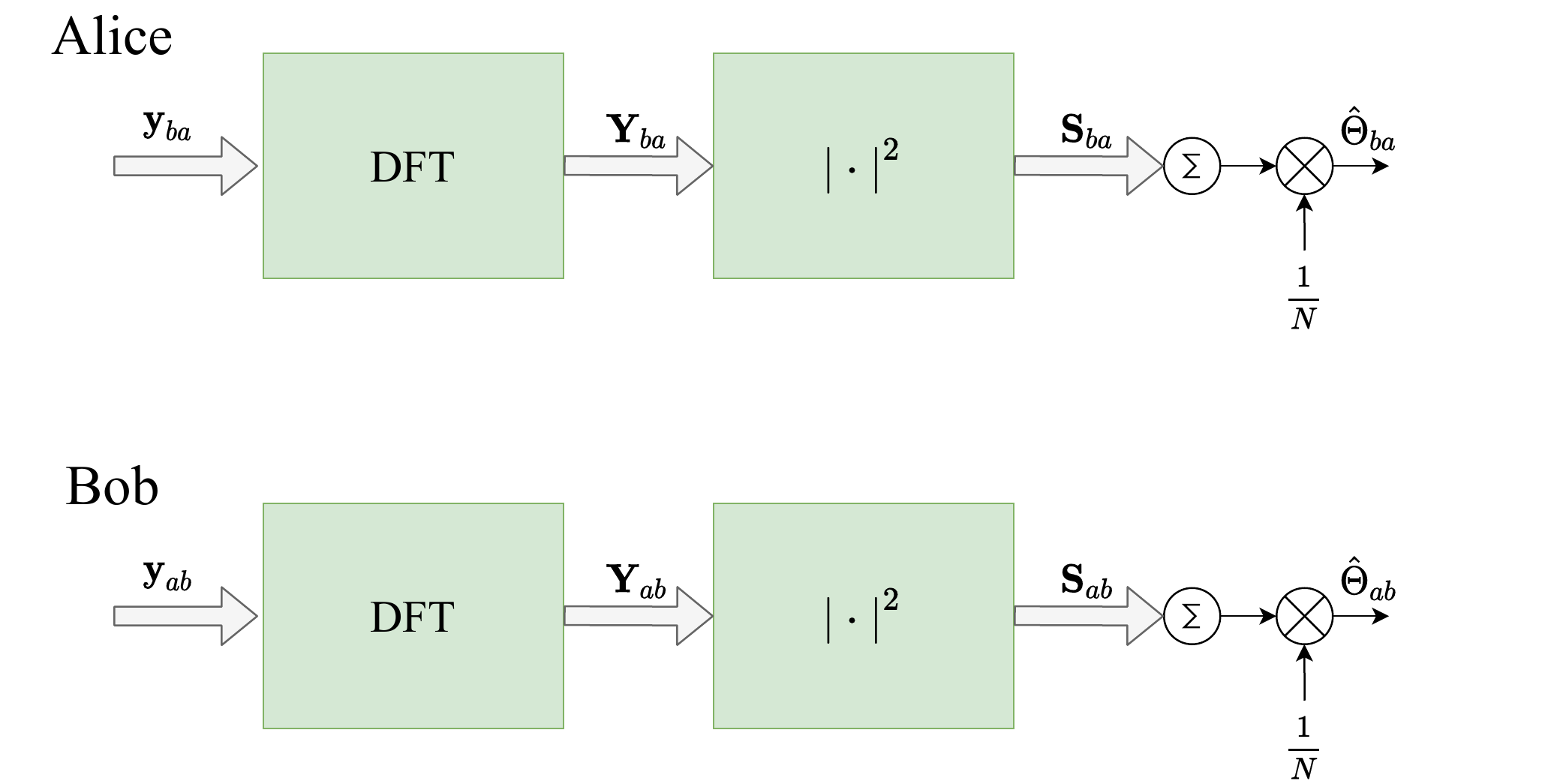}
	\caption{An illustration of NPSDS estimation at Alice and Bob.}
	\label{fig:psd}
\end{figure}

\textcolor{black}{In order to find the maximum value of the log-likelihood function, we evaluate the value of the function when its first derivative is zero:}

\begin{equation}
\frac{\partial \mathcal{L}(S_{jk}(i);{\Theta}_{jk})}{\partial \Theta_{jk}}= -\frac{N}{{\Theta}_{jk}}+\frac{\sum\limits_{i=1}^{N}S_{jk}(i)}{\Theta^2_{jk}}=0.
\end{equation}
The result of this equation provides us the ML estimator for the ${\Theta}_{jk}$ parameter as 
\begin{equation}
\hat{\Theta}_{jk}= M^{jk}_s= \frac{\sum\limits_{i=1}^{N}S_{jk}(i)}{N},
\end{equation}
where $ M^{jk}_s$ is the sample mean of the observed power spectral density samples. The steps followed in NPSDS estimation block can be summarized as in Figure \ref{fig:psd}.
\subsection{Quantization}
After obtaining  $\hat{\Theta}_{jk}$ values, terminals quantize and encode this value in order to obtain a secret key sequence. In this paper, we assume uniform quantizer with a step size $\Delta$ as in \cite{Topal}. From $N$ observations, we get $Q$ number of quantized secret key bits as $q_{b}= \left \lfloor \frac{\hat{\Theta}_{ab}}{\Delta} \right \rfloor$  and $q_{a}= \left \lfloor \frac{\hat{\Theta}_{ba}}{\Delta} \right \rfloor$, where $ \left \lfloor \cdot \right \rfloor$ denotes the floor function.

Note that, the main focus of this work is producing the raw key from a novel secrecy source, mainly the Doppler frequency shift. The key reconciliation and hybrid key generation algorithms in  \cite{Kurt_hybrid} can be applied to the quantized raw key bits in order to reduce the erroneous elements in the generated key bits.

\section{Key Disagreement Rate}

Key disagreement rate (KDR) is the ratio of mismatched bits in the generated keys. We adopt a similar approach to \cite{Topal} in order to obtain theoretical expressions of the KDR considering the proposed key generation system model.  To obtain the key disagreement rate, we normalize the estimated NPSDSs by multiplying with a normalization constant $ \eta=N/\Theta_{ab} $  as $\tilde{\Theta}_{ab}=\eta\hat{\Theta}_{ab}$ and $\tilde{\Theta}_{ba}=\eta\hat{\Theta}_{ba}$. Considering Proposition 1, we can state that $\Theta_{ba}=\Theta_{ab}=\Theta$. Assuming the uniform quantizer with a step size $\Delta$, an estimated $\tilde{\Theta}_{ab}$ is mapped to the $l^{\text{th}}$ quantization level, where $l= \left \lfloor \frac{\tilde{\Theta}_{ab}}{\Delta} \right \rfloor$ and the quantization interval is described by $I_l=[l\Delta, (l+1)\Delta]$. The probability that $\tilde{\Theta}_{ba}$ locates in  $I_l$ under given $\tilde{\Theta}_{ab}$ can be obtained by
\begin{equation}
P_l=\int\limits_{l\Delta}^{(l+1)\Delta}\rho(\tilde{\Theta}_{ba}|\tilde{\Theta}_{ab})d\tilde{\Theta}_{ba},
\end{equation}
where $\rho(\tilde{\Theta}_{ba}|\tilde{\Theta}_{ab})$ denotes the probability density function of $\tilde{\Theta}_{ba}$ given $\tilde{\Theta}_{ab}$. In order to obtain $\rho(\tilde{\Theta}_{ba}|\tilde{\Theta}_{ab})$, let us first obtain $\rho(\tilde{\Theta}_{ba}|\mathbf{Y}_{ab})$.
For a given $\mathbf{Y}_{ab}$, the observation signal at Alice can be modeled as $\mathbf{Y}_{ba}=\mathbf{Y}_{ab}-\mathbf{K}_{ab}+\mathbf{K}_{ba}$. In this conditional case,  $|{Y}_{ba}(i)| \sim \mathcal{R}(|Y_{ab}(i)|,\Theta)$, where $R\sim\mathcal{R}(\nu,\sigma^2)$ denotes a random variable following the Rician distribution for statistically independent $A\sim \mathcal{N}(\nu \cos(\alpha), \frac{\sigma^2}{2})$ and $B\sim \mathcal{N}(\nu \sin(\alpha), \frac{\sigma^2}{2})$ and $R=\sqrt{A^2+B^2}$. The normalized estimated NPSDS at Alice can be modeled as
\begin{equation}
\tilde{\Theta}_{ba}= \sum_{i=1}^{N} Z^2(i)= \frac{\sum_{i=1}^{N}\eta|Y_{ba}(i)|^2}{N},
\end{equation}
where $Z(i)\sim \mathcal{R}(|{Y}_{ab}(i)|,1)$. In this case, we can state that $\tilde{\Theta}_{ba} \sim \chi'^{2}_{k_a}(\lambda_a) $, where $\chi'^{2}_k(\lambda)$ denotes non-central chi-square distribution with $k$ degrees of freedom and $\lambda$ noncentrality parameter. The degrees of freedom can be given as $k_a=2N$. The noncentrality parameter is obtained by
\begin{equation}
\lambda_a=\sum_{i=1}^{N}\frac{\eta|{Y}_{ab}(i)|^2}{N} =\eta \hat{\Theta}_{ab}= \tilde{\Theta}_{ab}.
\end{equation}
As (13) shows the pdf of $\tilde{\Theta}_{ba}$ is parametrized by $N$ and $\tilde{\Theta}_{ab}$. Therefore $P_l$ becomes
\begin{equation}
P_l=\!\!\!\!\!\!\int\limits_{l\Delta}^{(l+1)\Delta} \frac{1}{2} e^{-\frac{(x+\tilde{\Theta}_{ab})}{2}}\left(\frac{x}{\tilde{\Theta}_{ab}}\right)^{ \frac{(N-1)}{2}}\!\!\!\!\!\!\!\!\!\!\!\! I_{2N / 2-1}\left(\sqrt{\tilde{\Theta}_{ab} x}\right)  dx,
\end{equation}
where $I_v(y)$ denotes the modified Bessel function of the first kind. The closed form expression for the $P_l$ can be expressed by 
\begin{equation}
P_l= Q_N\left(\sqrt{\tilde{\Theta}_{ab}}, \sqrt{l \Delta }\right)-Q_N\left(\sqrt{\tilde{\Theta}_{ab}}, \sqrt{(l+1) \Delta }\right),
\end{equation}
where $Q_N(\alpha,\beta)$ denotes the Marcum-Q fuction. Considering that $\tilde{\Theta}_{ab}$ is also a random variable, the key matching probability can be obtained by
\begin{equation}
P_c=\int_{0}^{\infty}P_l \rho(\tilde{\Theta}_{ab}) d\tilde{\Theta}_{ab}.
\end{equation} 
Considering $\tilde{\Theta}_{ab}$ is a random variable with Gamma distribution with $N$ shape parameter and $1$ scale parameter, where its pdf can be described by
\begin{equation}
 \rho(\tilde{\Theta}_{ab})=\frac{1}{\Gamma(N)}\tilde{\Theta}^{N-1}_{ab}e^{-\tilde{\Theta}_{ab}}.
\end{equation}

In the expression above, $\Gamma(N)$ denotes the Gamma function, where $\Gamma(N)=(N-1)!$ for all positive integer values of $N$. The resulting $P_c$ becomes
\begin{equation}
\begin{aligned}
P_c&= \frac{1}{\Gamma(N)}\int_{0}^{\infty} Q_N\left(\sqrt{\tilde{\Theta}_{ab}}, \sqrt{l \Delta }\right)\tilde{\Theta}^{N-1}_{ab}e^{-\tilde{\Theta}_{ab}}d\tilde{\Theta}_{ab}\\
&-\frac{1}{\Gamma(N)}\int_{0}^{\infty}Q_N\left(\sqrt{\tilde{\Theta}_{ab}}, \sqrt{(l+1) \Delta }\right)  \tilde{\Theta}^{N-1}_{ab}e^{-\tilde{\Theta}_{ab}} d\tilde{\Theta}_{ab}
\end{aligned}.
\end{equation}

In order to obtain a closed form approximation for $P_c$, we use generalized Gauss-Laguerre quadrature as $\int_{0}^{\infty} \psi^ae^{-\psi} f(\psi) d\psi \approx \sum\limits_{m=1}^{M}w_mf(\psi_m)$, where $$w_m= \frac{\Gamma(M+a+1)\psi_m}{M!(M+1)^2\left(L_{M-1}^{(a)}(\psi_m)\right)^2}.$$
The degree of the polynomial is denoted by $M$, and as $M\rightarrow \infty$, the approximation becomes equality\cite{laguerre}. $\psi_m$ denotes the $m^{\text{th}}$ root of the polynomial $L_{M-1}^{(a)}$. After applying the generalized Gauss-Laguerre approximation, $P_c$ is approximated by $\tilde{P}_c$ as
 \begin{equation}
 \text{\small $\tilde{P}_c= \Upsilon  \sum\limits_{m=1}^{M} \xi_m \left[Q_N\left(\sqrt{\psi_m}, \sqrt{l \Delta }\right)\right. 
 	-\left. Q_N\left(\sqrt{\psi_m}, \sqrt{(l+1) \Delta }\right)\right],$}
\end{equation}
where $\Upsilon=\frac{\Gamma(M+N)}{\Gamma(N)M!(M+1)^2}$ and $\xi_m=\frac{\psi_m}{\left(L_{M-1}^{(a)}(\psi_m)\right)^2}$. Resulting KDR can be approximated by 
\begin{equation}
\text{KDR} \approx 1-\tilde{P}_c.
\end{equation}
By adapting $M$, the approximation to the exact KDR would be tighter. In order to highlight the tightness of the approximation, in the following section, the obtained KDR expression in (19)-(20) is compared with the simulations. 
\begin{figure*}[t]
	\begin{center}	
		\begin{adjustbox}{max width=\textwidth}	
			\subfigure[]{
				\label{fig:pdfN10}
				\includegraphics[width=\textwidth]{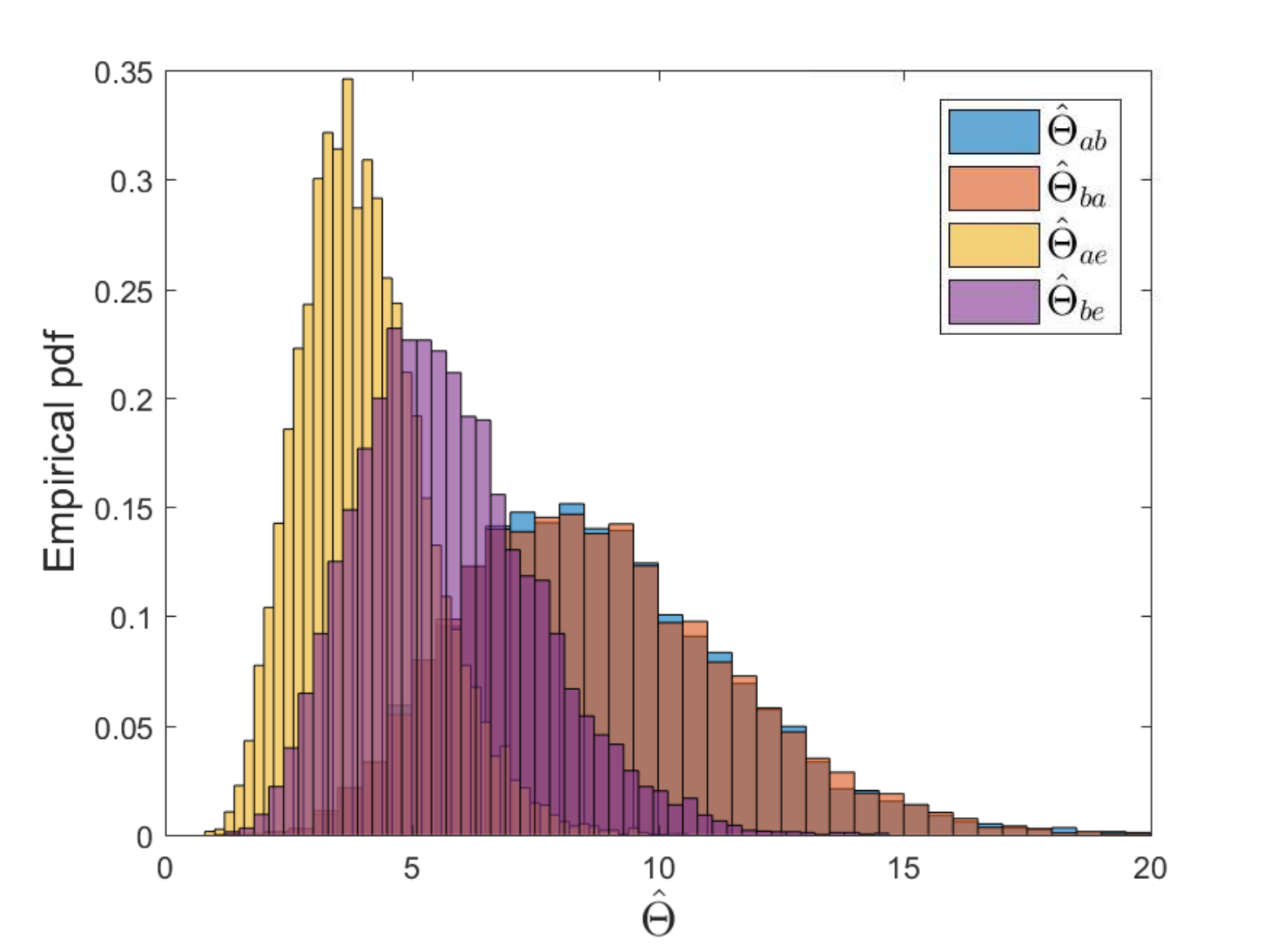} }
\hfill

			\subfigure[]{
				\label{fig:pdfN20}
				\includegraphics[width=\textwidth]{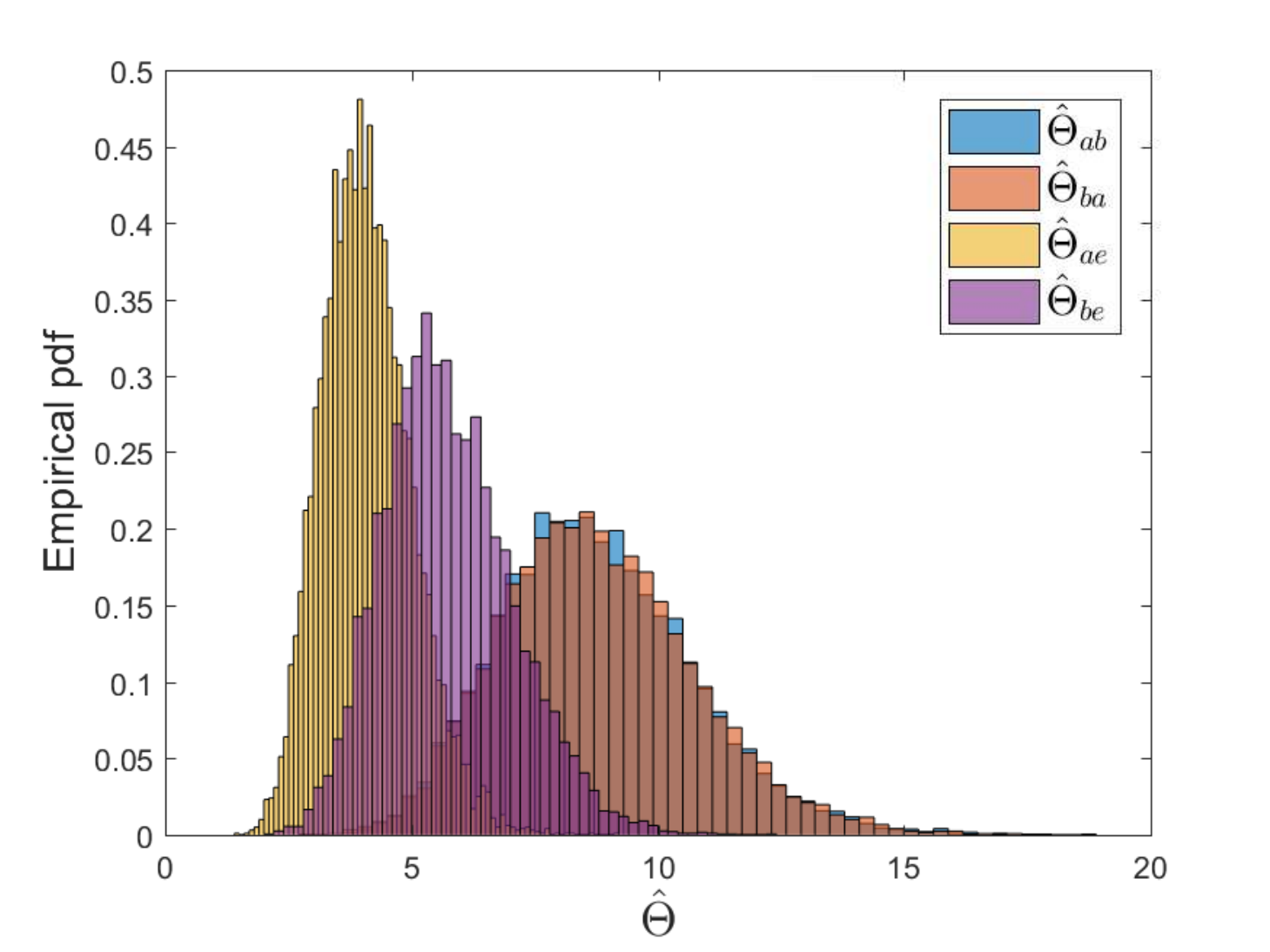} }
\hfill
			\subfigure[]{
				\label{fig:pdfN50}
				\includegraphics[width=\textwidth]{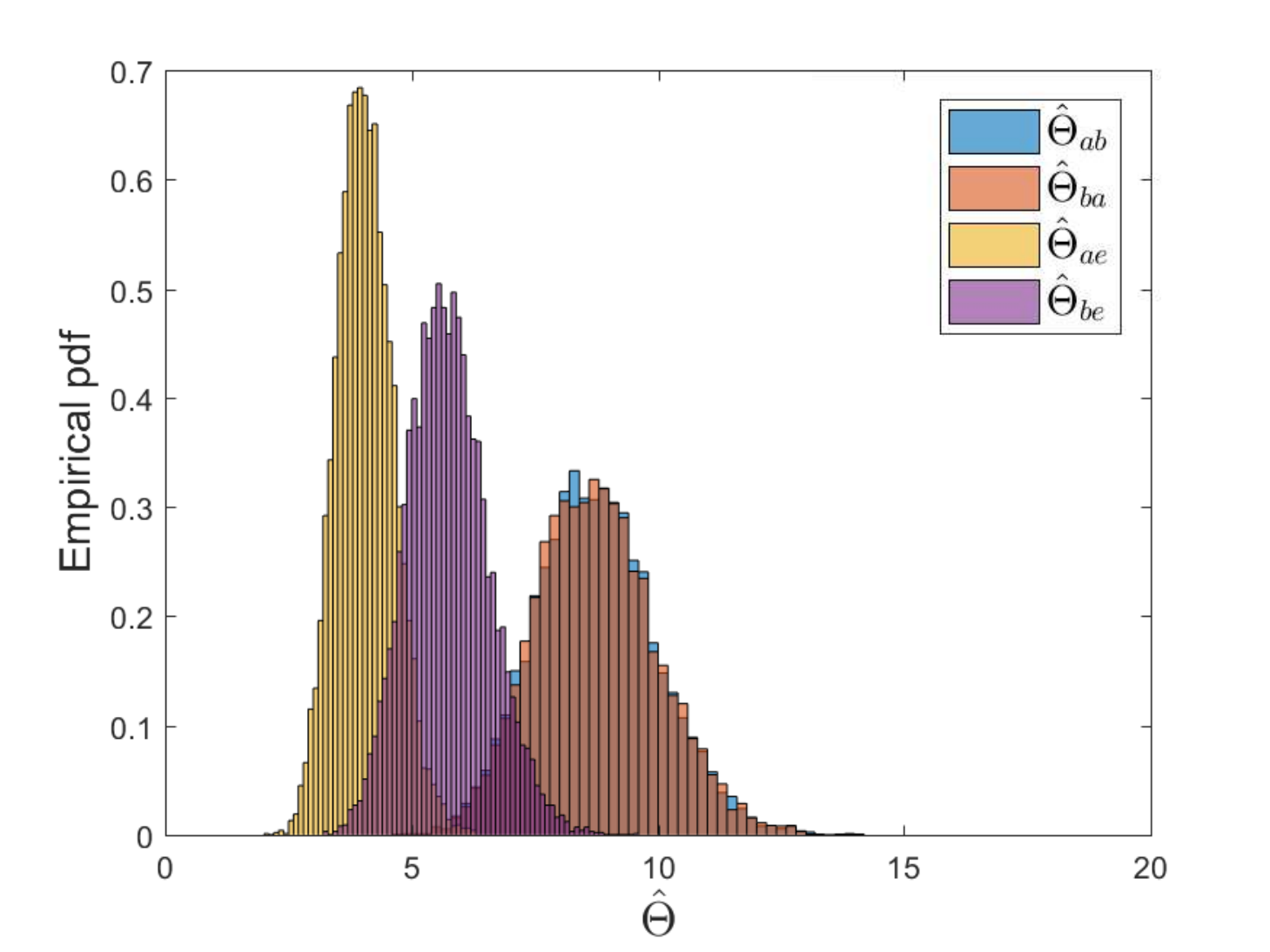} }
		\end{adjustbox}
	\end{center}
	\vspace{-0.4cm}
	\caption{Empirical probability density functions for estimated NPSDSs for three different $N$ values, (a) $N=10$, (b) $N=20$ and (c) $N=50$.}
	\label{fig:MixHist-sc1}
\end{figure*}

\section{Numerical Analysis}
We consider three {\color{black}spacecraft}s Alice, Bob and Eve. Alice and Bob employ the proposed key generation scheme in Section II in order to obtain a secret key. In a single key duration, first Alice and Bob respectively transmits $N$ pilot symbols. Then, they estimate NPSDSs, and feed them into uniform quantizer. In the meantime, Eve observes the transmitted pilot symbols from both {\color{black}spacecraft}s, and obtain NPSDSs for each transmission. {\color{black} As the worst case, we assume that Eve has a priori knowledge of the NPSDS of the pilot message.} Each of the legitimate nodes generate single secret key, while Eve generates two different versions of the secret key. Numerically KDR in a single key duration can be described by
\begin{equation}
KDR_{jk}= \begin{cases}
1, q_j\neq q_k \\ 0, q_j=q_k
\end{cases}.
\end{equation}
Considering $D$ number of key durations, KDR for {\color{black}spacecraft}s $k$ and $j$ becomes
$$
\text{KDR}_{jk}= \frac{\sum\limits_{t=1}^{D}KDR_{jk}(t)}{D}.
$$
 The simulation parameters are given in Table \ref{tab:simulation_parameters}, where the values are obtained from \cite{numerical}. The NPSDS of BPSK simulation is utilized as in \cite{proakis}. {\color{black} Note that, as the mobility of two nodes continiues to change over time as in the {\color{black}spacecraft}s, the proposed key generation mechanism can also be applied at higher frequency bandwidths.}

\begin{table}[t]
	\caption{Simulation parameters}
	\centering
	\begin{tabular}{|l|l|}
		\hline
		Carrier frequency  & 1 GHz   \\ \hline
		$\omega_{ab}$      & 200 MHz \\ \hline
		$\omega_{ae}$      & 500 MHz \\ \hline
		$\omega_{be}$      & 400 MHz \\ \hline
		Symbol energy      & 10 dB   \\ \hline
		Modulation type    & BPSK    \\ \hline
		Noise variance     & 1 dB    \\ \hline
		Path loss exponent & 2       \\ \hline
	\end{tabular}
	\label{tab:simulation_parameters}
\end{table}
\begin{figure}[h]
	\centering
	\includegraphics[width=0.95\linewidth]{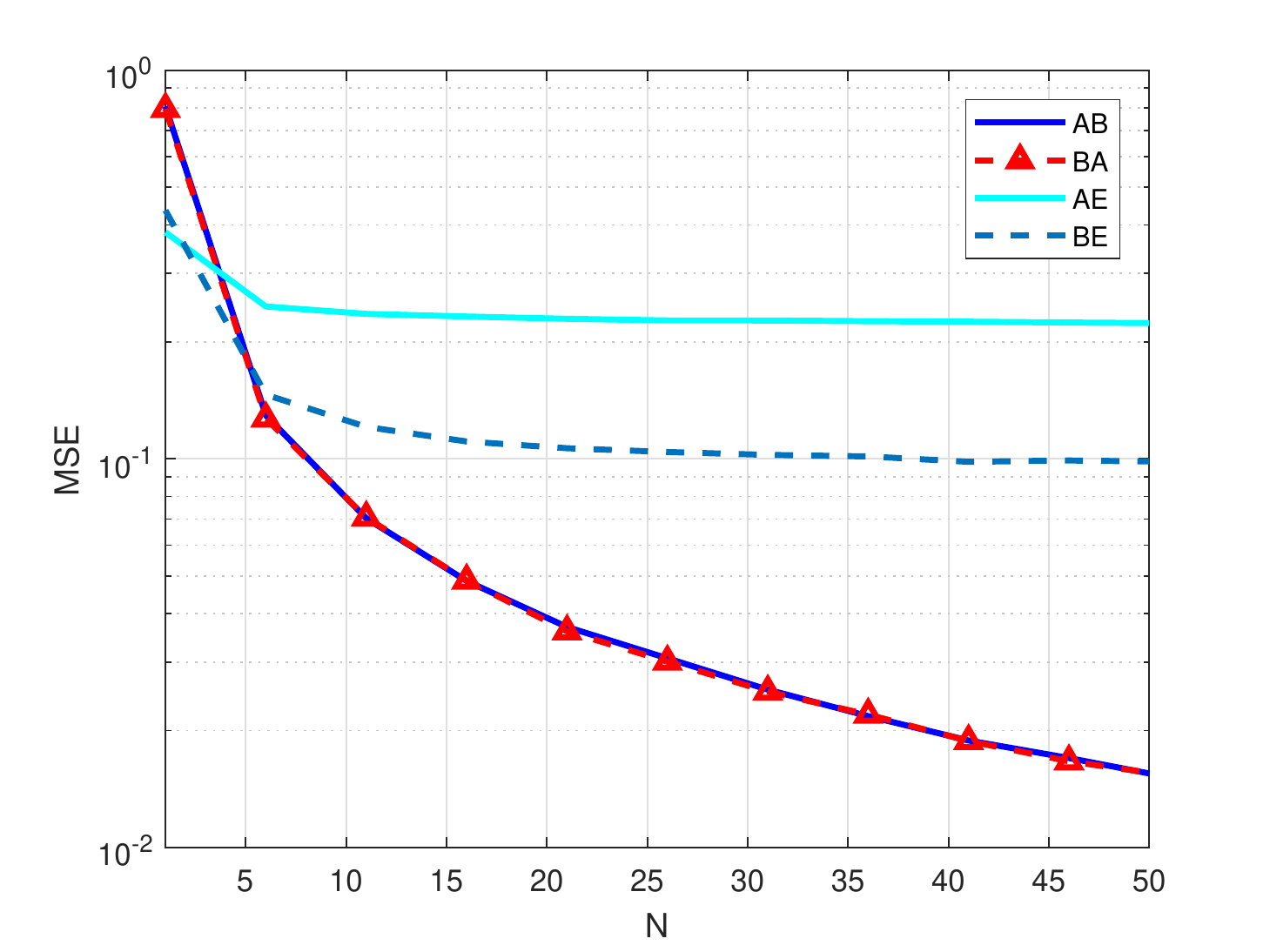}
	\caption{MSE vs. channel observations considering the different ISLs.}
	\label{fig:MSE}
\end{figure}
Figure \ref{fig:MixHist-sc1} shows the empirical probability distributions (pdfs) of the estimated NPSDSs at Alice, Bob and Eve. Since Eve captures both the transmission from Alice to Bob and from Bob to Alice, she obtains two different versions of the estimated NPSDSs. From left to right (Fig. \ref{fig:pdfN10}-\ref{fig:pdfN20}-\ref{fig:pdfN50}), the number of observations increases. One remark is that as the number of observation increases, the variance of the pdfs become smaller. The distribution at Alice and Bob are almost identical as defined in (4) and (5). On the other hand, the pdfs at Eve diverges from Alice and Bob, since their relative velocities and Doppler frequencies are different. 

In order to focus on the relationship between the number of observations and the estimation disparities, we utilize mean squared error (MSE) as in Figure \ref{fig:MSE}. The MSE for the NPSDS  estimator for the $j-k$ link can be given as $$\text{MSE}_{jk}=\frac{1}{N} \sum_{i=0}^{N}|\hat{\Theta}_{jk}-\Theta_{jk}|^2.$$ As indicated in the figure, the estimations at Alice and Bob converges as the number of observations increases. Even though estimation error at Eve also decreases, the error becomes stable after $N=5$. Since the relative velocity of Alice and Bob differs from the relative velocity of Alice and Eve, and the relative velocity of Bob and Eve, the Doppler frequency values observed at Eve is also different than Alice and Bob. The results in both Figure \ref{fig:MixHist-sc1} and Figure  \ref{fig:MSE} show that the NPSDSs or Doppler frequencies can be utilized as a secret source between {\color{black} two separately moving nodes.}

\begin{figure}[tbh]
	\centering
	\includegraphics[width=0.95\linewidth]{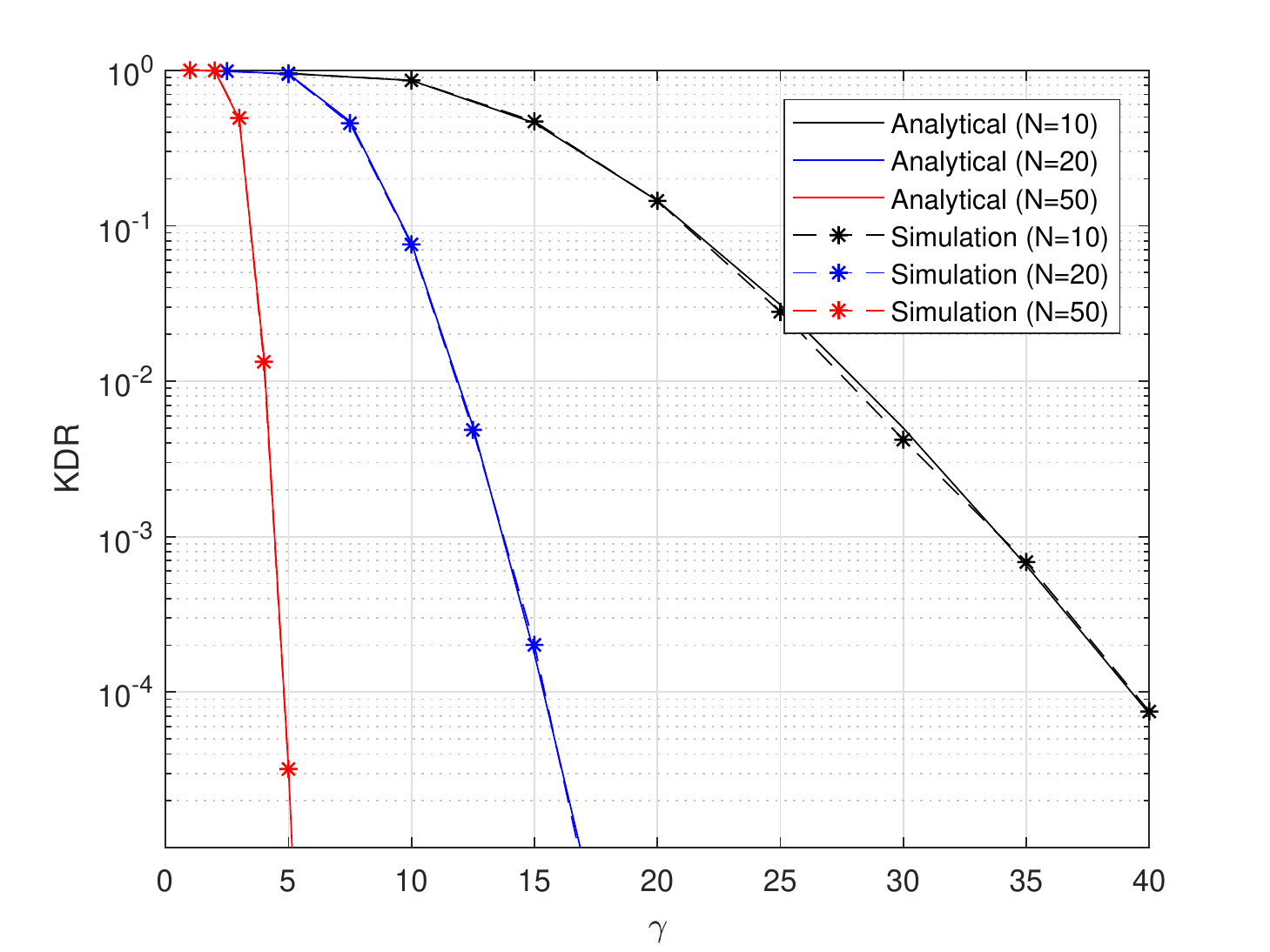}
	\caption{KDR vs. normalized quantization interval considering different number of channel taps.}
	\label{fig:KDR}
\end{figure}

Figure \ref{fig:KDR} provides KDR values of the generated keys at Alice and Bob for different normalized quantization intervals, where $\gamma=\frac{\Delta}{N}$. As the quantization interval increases, KDR decreases. Since as the number of observation at Alice and Bob increases, the error at their estimates decreases. Therefore, increment in the number of observation samples decreases KDR more dramatically. Note that, increment in the number of observation samples would also require transmitting more pilot symbols and reduces the efficiency of the protocol. Solid lines show the approximate KDR by {\color{black} Eqs. (19)-(20)} while the stars indicate the simulation results. Note that, $M=100$ is selected in (19). The tightness of the proposed approximations can be observed from the figures.

\section{Conclusion}
In this work, we have proposed a security mechanism for the inter-{\color{black}spacecraft} links (ISLs) first time in the literature. The proposed mechanism ensures continuous secrecy between two distant nodes. The secrecy of the proposed method is based on the symmetric Doppler frequency measurements of the {\color{black}spacecrafts}. Theoretical expressions of the key disagreement rate (KDR) are derived considering the estimation errors at {\color{black}spacecraft}. Tight approximations to KDR expressions are obtained by using Marcum-Q functions and generalized Gauss-Laguerre quadrature (GLQ). The provided numerical results highlight the tightness of the given approximations, and indicate the applicability of the proposed key generation mechanism. As a future work, we consider a hybrid key generation mechanism that harness different secrecy resources of the physical layer (channel fading, Doppler frequency, RSS) that compensates the performance of different mechanisms.

\bibliographystyle{IEEEtran}

\input{Doppler_space_key.bbl}

\end{document}